\documentclass[prd,aps,floats,nofootinbib,11pt]{revtex4}
\usepackage{amsmath}
\usepackage{amssymb}

\numberwithin{equation}{section}

\usepackage{epsfig}

\def\beq{\begin{equation}}
\def\eeq{\end{equation}}
\def\ber{\begin{eqnarray}}
\def\eer{\end{eqnarray}}
\def\l{\Lambda}

\def\apj{{Astroph.\@ J.\ }}

\def\prd{{Phys.\@ Rev.\@ D\ }}
\def\pd{{Phys.\@ Rev.\@ D\ }}
\def\cqg{{Class. Quant. Grav.\ }}

\def\plb {{Phys.\@ Lett.\@ B\ }}

\def\etal{{\it et al.}}

\def\b{{\rm b}}
\def\m{{\rm m}}
\def\l{\Lambda}

\begin{document}
\sloppy

\title{Cosmic Acceleration and Extra Dimensions}

\author{Varun Sahni}\email{varun@iucaa.ernet.in}
\affiliation{Inter-University Centre for Astronomy \& Astrophysics, Post Bag 4,
Pune~411~007, India}
\author{Yuri Shtanov}\email{shtanov@bitp.kiev.ua}
\affiliation{Bogolyubov Institute for Theoretical Physics, Kiev 03680, Ukraine}

\thispagestyle{empty}

%\maketitle
\sloppy

\begin{abstract}
Brane cosmology presents many interesting possibilities including: phantom acceleration
($w<-1$), self-acceleration, %(DGP model),
unification of dark energy with inflation, transient acceleration, loitering cosmology,
new singularities at which the Hubble parameter remains finite, cosmic mimicry, etc. The
existence of a {\em time-like\/} extra dimension can %within the Randall--Sundrum setting,
result in a singularity-free {\em cyclic\/} cosmology.
\end{abstract}

\maketitle

%\pacs{PACS number(s): 04.50.+h, 98.80.Hw}

\bigskip

{\small \em It gives us great pleasure to write this paper on the occasion of Sergei
Odintsov's fiftieth birthday. Sergei has written many excellent papers over the past
several decades, and we hope that he will write an equal number in the coming 50
years\,!}

\section{Introduction}

Considerable evidence points to a universe that is accelerating \cite{data}. Although the
cosmological constant $\Lambda$ provides conceptually the simplest explanation of cosmic
acceleration, its enigmatically small value has led researchers to explore alternative
avenues for generating an accelerating universe \cite{DE_review,ss06,modified_grav}. In
this paper, we shall confine our attention to brane cosmology described by the fairly
general action \cite{CH,Shtanov1}
\begin{equation} \label{action}
S = M^3 \left[\int_{\rm bulk} \left( {\cal R} - 2 \Lambda_{\rm b}
\right) - 2 \int_{\rm brane} K \right] + \int_{\rm brane} \left( m^2 R - 2
\sigma \right) + \int_{\rm brane} L \left( h_{ab}, \phi \right) \, .
\end{equation}
Here, ${\cal R}$ is the scalar curvature of the metric $g_{ab}$ in the five-dimensional
bulk, and $R$ is the scalar curvature of the induced metric $h_{ab}$ on the brane. The
brane is considered to be a boundary of the bulk space, $K$ is the trace of the extrinsic
curvature tensor of the brane, and $L (h_{ab}, \phi)$ denotes the Lagrangian density of
the four-dimensional matter fields $\phi$ whose dynamics is restricted to the brane. $M$
and $m$ denote, respectively, the five-dimensional and four-dimensional Planck masses,
$\Lambda_{\rm b}$ is the five-dimensional (bulk) cosmological constant, and $\sigma$ is
the brane tension.

Action (\ref{action}) leads to the following cosmological evolution equation on the brane
\cite{CH,Shtanov1,ss02}:
\begin{equation} \label{cosmo}
m^4 \left( H^2 + \frac{\kappa}{a^2} - \frac{\rho + \sigma}{3 m^2} \right)^2 =
\epsilon M^6  \left(H^2 + \frac{\kappa}{a^2} - \frac{\Lambda_{\rm b}}{6} -
\frac{C}{a^4} \right) \, ,
\end{equation}
where $\epsilon = 1$ if the extra dimension is space-like, and $\epsilon = -1$ if it is
time-like, $C$ is an integration constant reflecting the presence of a black hole in the
bulk space, the term $C / a^4$ is usually called `dark radiation,' and $\kappa = 0, \pm
1$ reflects the spatial curvature of the brane.

Several important cosmological scenarios arise
 as special
cases of (\ref{cosmo}), including:
\begin{enumerate}
\item General Relativity ($M=0$, $\Lambda_{\rm b}=0$),
\item The self-accelerating Dvali--Gabadadze--Porrati (DGP) brane \cite{DGP}
    ($\Lambda_{\rm b}=0$, $\sigma = 0$),
\item The Randall--Sundrum (RS) model \cite{RS} ($m=0$).
\end{enumerate}

Indeed, action (\ref{action}) can result in cosmological models which differ from GR
either {\em early on\/} or at {\em late times\/}. The Randall--Sundrum model belongs to
the former class whereas the DGP brane is a famous example of the latter category. Other
interesting properties of models with late-time acceleration include phantom expansion
\cite{ss02}, loitering \cite{loiter} and cosmic mimicry \cite{mimicry}, all of which
shall be briefly discussed in this paper.

\section{Unified models of inflation and dark energy}

An intriguing question faced by cosmologists is why the universe accelerates twice:
during inflation and again at the present epoch. The notion of quintessential inflation
--- attempting to unify early and late acceleration --- was originally suggested in the
context of GR by Peebles and Vilenkin \cite{pv99}. The possibility that braneworld models
could provide a more efficient realisation of this scenario was discussed in
\cite{copeland,sami,quint-inf1}. Note that $m = 0$ in the Randall--Sundrum model, so that
(\ref{cosmo}) reduces to
\begin{equation}
H^2 + \frac{\kappa}{a^2} = \frac{8\pi G}{3} \left( \rho +
\frac{\rho^2}{2\sigma} \right) + \frac{\Lambda}{3} + \frac{C}{a^4} \, ,
\label{eq:brane}
\end{equation}
where $G = {\epsilon \sigma}/{12 \pi M^6 }$, $\Lambda = {\Lambda_{\rm b}}/{2} + {\epsilon
\sigma^2}/{3 M^6} $. A scalar field evolving on the brane satisfies the usual equation
\begin{equation}
{\ddot \phi} + 3H {\dot \phi} + V'(\phi) = 0\, ,
\label{eq:kg}
\end{equation}
where $H$ is given by (\ref{eq:brane}), %with $\rho = \half{\dot\phi}^2 +
%V(\phi)$.
and the energy density and pressure of the scalar field are, respectively,
\begin{equation}
\rho_\phi = \frac12 \dot{\phi}^2 + V (\phi) \, , \qquad  P_\phi =
\frac12 \dot{\phi}^2 - V (\phi) \, . \label{eq:6}
\end{equation}
If $\sigma > 0$, then the new term $\rho^2/2\sigma$ in (\ref{eq:brane}) {\em increases\/}
the damping experienced by the scalar field as it rolls down its potential, making the
inflationary condition $P \simeq -\rho$ easier to achieve.

Consequently, inflation can be driven by {\em steep potentials\/}, such as $V \propto
\phi^{-\alpha}$, $\alpha > 1$, which are usually associated with dark energy (DE)
\cite{copeland}\,. Thus the class of potentials giving rise to inflation increases and
the possibility of realising inflation becomes easier in brane cosmology
\cite{copeland,sami,quint-inf1}. Brane inflation leaves behind an imprint on the
cosmological gravity-wave background by increasing its amplitude and creating a distinct
`blue tilt' in its spectrum, thereby permitting verification through future LISA-type
searches \cite{sami}.

\section{Cyclic cosmology on the brane}

If, in the Randall--Sundrum model, the extra dimension is {\em time-like\/}, then the
big-bang singularity is completely absent\,! To see this, consider equation
(\ref{cosmo}), this time with $\epsilon = -1$. The resulting braneworld dynamics is
described by \cite{bounce}
\begin{equation}
H^2 + \frac{\kappa}{a^2} = \frac{8\pi G}{3} \left( \rho -
\frac{\rho^2}{2|\sigma|} \right)~,
\label{eq:bounce}
\end{equation}
where we have ignored the contribution from $\Lambda$ and dark radiation. Since the `$+$'
sign within the bracket in (\ref{cosmo}) is replaced by a `$-$' sign, this braneworld
model can be regarded as dual to the RS model \cite{copeland04}. Consequently, $H = 0$
when $\rho_{\rm bounce} = 2|\sigma|$, i.e., the universe {\em bounces\/} when the density
of matter has reached a sufficiently large value. Note that the singularity-free nature
of the early universe is {\em generic\/} and does not depend upon whether or not matter
violates the energy conditions \cite{loop}.

\begin{figure*}[ht]
\centerline{ \psfig{figure=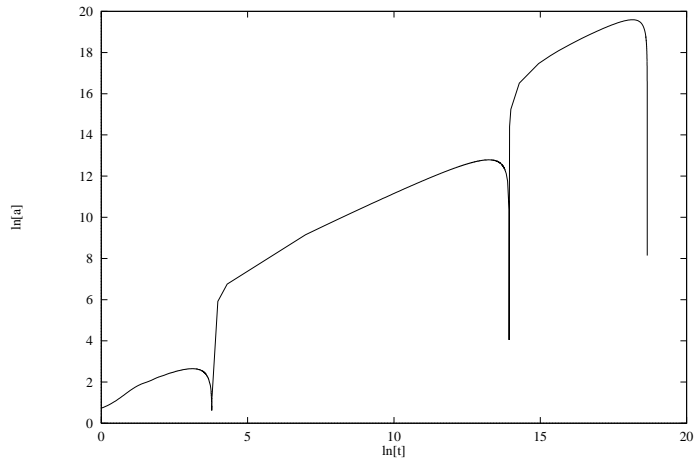,width=0.40\textwidth,angle=-90} }
\bigskip
\caption{\small A cyclic universe with increasing successive expansion
maxima. Figure courtesy of \cite{kanekar}.
}
\label{fig:cyclic}
\end{figure*}

This scenario has also been used to construct cyclic models of the universe
\cite{bfk04,kanekar,piao04a}. For instance, in \cite{bfk04} dark energy is postulated to
be a phantom having $w< -1$. Consequently, its energy density {\em grows as the universe
expands}, $\rho_{\rm ph} \propto a^{-3(1+w)}$, while the density of normal
matter/radiation decreases. Since $\rho$ grows at small {\em as well as} large values of
the expansion factor, the Hubble parameter passes through zero twice: (i) at early times
when the bounce in (\ref{eq:bounce}) is caused by the large radiation density and (ii)
during late times, when the large value of the phantom density leads to $H=0$ in
(\ref{eq:bounce}) and initiates the universe's recollapse. The universe can also
recollapse if it is spatially closed or if dark energy falls to negative values, as in
the case of DE with a cosine potential \cite{cos}. In the braneworld context, such models
will also be `cyclic' in the sense that they will pass through an infinite number of
nonsingular expanding-contracting epochs \cite{kanekar,piao04a}.

%\section{Braneworld dark energy}
\section{Phantom brane}

The previous examples showed how brane cosmology could differ from that in GR at
{\em early times}. We now demonstrate that the same can happen at late times.

The cosmological equation (\ref{cosmo}) can be expressed in the following way in a
spatially flat braneworld \cite{ss02}:
\ber\label{hubble}
H^2 (a)  &=& \frac{A}{a^3} + \Lambda_{\rm eff} \, , \\
%The term $\Lambda_{\rm eff}$ is composed of two terms, namely, a constant
%$\Lambda$-term and a `screening term':
\label{eq:screening} \Lambda_{\rm eff} &=& \left(B + \frac{2}{\ell^2} \right) \pm
\frac{2}{\ell^2} \sqrt{1 + \ell^2 \left( \frac{A}{a^3} + B  - \frac{\Lambda_{\rm b}}{6} -
\frac{C}{a^4} \right)} \, ,
\eer
%$$
%\hskip 0.1cm \Downarrow \hskip 2.9cm \Downarrow
%$$
%\hskip 6.5cm {\Large{$\Lambda$}} \hskip 1.4cm Screening term
%\bigskip
where
\begin{equation}\label{ab}
A = \frac{\rho_{0} a_0^3}{3 m^2} \, , \quad B = \frac{\sigma}{3 m^2} \, , \quad
\ell = \frac{2 m^2}{M^3} \, .
\end{equation}
The two branches of solutions, designated by the $\pm$ sign in (\ref{eq:screening}),
correspond to two possible ways of embedding the brane in the bulk space
\cite{CH,ss02,Deffayet}.

Consider the branch with the `$-$' sign, which we allude to as Brane\,1. Since, in this
case, the second term in (\ref{eq:screening}) decreases with time, the value of the
effective cosmological constant $\Lambda_{\rm eff}$ {\em increases\/}
\cite{ss02,lue-starkman}. Therefore, the braneworld expansion proceeds as that of a
universe which is described by general relativity and filled with phantom ($w_{\rm eff} <
-1$), but, unlike phantom, matter on the brane does not violate the weak energy condition
$\rho+P\geq 0$. From (\ref{eq:screening}) it is also clear that the universe evolves to
$\Lambda$CDM in the future and does not encounter a {\em Big Rip\/} singularity peculiar
to phantom DE
\cite{caldwell02}. 
The fact that the braneworld model (\ref{hubble}), (\ref{eq:screening}) can give rise to
phantom-like behaviour can also be seen if we rewrite it in terms of the cosmological
redshift $z$, neglecting the dark-radiation term $C/a^4$, so that \cite{ss02,loiter}
\beq \label{hubble1}
\frac{H^2(z)}{H_0^2} = \Omega_{\rm m} (1\!+\!z)^3 + \Omega_\sigma + 2 \Omega_\ell \pm 2
\sqrt{\Omega_\ell}\, \sqrt{\Omega_{\rm m} (1\!+\!z )^3 + \Omega_\sigma + \Omega_\ell +
\Omega_{\Lambda_{\rm b}}} \, ,
\eeq
where
\beq \label{omegas}
\Omega_{\rm m} =  \frac{\rho_0}{3 m^2 H_0^2} \, , \quad \Omega_\sigma = \frac{\sigma}{3
m^2 H_0^2} \, , \quad \Omega_\ell = \frac{1}{\ell^2 H_0^2} \, , \quad
\Omega_{\Lambda_{\rm b}} = - \frac{\Lambda_{\rm b}}{6 H_0^2} \, .
\eeq
The current value of the effective equation of state is given by \cite{ss02,loiter}
\begin{equation}
w_{\rm eff} = \frac{2 q_0 - 1}{3 \left( 1 - \Omega_{\rm m} \right)} = - 1 \pm
\frac{\Omega_{\rm m}}{1 - \Omega_{\rm m}} \, \frac{\displaystyle \sqrt{\Omega_\ell}}
{\displaystyle \left|\sqrt{1 + \Omega_{\Lambda_{\rm b}}} \mp \sqrt{\Omega_\ell} \right|} \, ,
\label{eq:brane_state}
\end{equation}
from which we see that $w_{\rm eff} \leq -1$ for Brane\,1, described by the lower sign
option in (\ref{eq:brane_state}). (The second choice of embedding, Brane\,2, gives
$w_{\rm eff} \geq -1$.) It is important to note that all Brane\,1 models have $w_{\rm
eff} \leq -1$ and $w(z) \simeq -0.5$ at $z \gg 1$ and successfully cross the `phantom
divide' at $w = -1$ \cite{brane_obs1,brane_obs2}.

Note that the DE equation of state in modified gravity models is notional and not
physical \cite{ss06} and for this reason the statefinder and the $Om$ diagnostic
\cite{statefinder,om} provide a more comprehensive picture of cosmic acceleration in such
models.

\section{\bf The DGP model}

A very interesting braneworld model, suggested by Dvali, Gabadadze and Poratti
\cite{DGP,Deffayet,DGP1}, is based on action (\ref{action}) with $\Lambda_{\rm b}=0$,
$\sigma = 0$. The spatially flat {\em self-accelerating\/} DGP brane without dark
radiation is described by
\beq\label{eq:dgp}
H = \sqrt{\frac{8\pi G \rho_{\rm m}}{3} + \frac{1}{\ell^2}} + \frac{1}{\ell}~.
\eeq
In this case, the universe accelerates because gravity becomes five-dimensional on length
scales $R > \ell = 2H_0^{-1}(1-\Omega_{\rm m})^{-1}$. Comparing (\ref{eq:dgp}) with the
corresponding expression for LCDM
\beq \label{h-lcdm}
H = \sqrt{\frac{8\pi G \rho_{\rm m}}{3} + \frac{\Lambda}{3}}~,
\eeq
we find that
the cosmological constant $\Lambda$ is replaced by a new fundamental constant
$M$ in the DGP model. However, unlike $\Lambda$, the value of $M$ is not
 unnaturally small. Indeed,
$M \sim 10$\,MeV can give rise to a universe which accelerates today with $\Omega_{\rm m}
\simeq 0.3$ \cite{DGP,Deffayet,DGP1}.

While providing an interesting alternative to dark energy, the DGP model does not agree
with observations as well as LCDM \cite{goobar,brane_obs1,maartens_DGP}. But the biggest
stumbling block for this model appears to be theoretical and has to do with the existence
of a ghost on the self-accelerating branch of solutions, which poses grave difficulties
for the DGP gravity; see \cite{ghost} and references therein.

\section{\bf Quiescent Cosmological Singularities}

A new feature of brane cosmology described by (\ref{hubble1}) is the presence of
singularities at which the density, pressure and Hubble parameter remain finite while the
deceleration parameter diverges \cite{quiescent}. Then these {\em quiescent\/}
singularities arise when the inequality
\begin{equation} \label{constraint}
\Omega_\sigma + \Omega_\ell + \Omega_{\Lambda_{\rm b}} \equiv
\left( \sqrt{1 + \Omega_{\Lambda_{\rm b}}} \mp \sqrt{\Omega_\ell} \right)^2
- \Omega_{\rm m} < 0 \, ,
\end{equation}
is satisfied, in which case the expression under the square root of (\ref{hubble1})
becomes zero at a suitably late time and the cosmological solution {\em cannot be
extended beyond this point}. (The quiescent singularity is the result of a singular
embedding of the brane in the bulk \cite{quiescent}.)

The limiting redshift, $z_s = a_0/a(z_s) - 1$, at which the braneworld becomes singular
is given by
\begin{equation}
z_s = \left [1 -\frac{\displaystyle \left( \sqrt{1 + \Omega_{\Lambda_{\rm b}}}
\mp \sqrt{\Omega_\ell} \right)^2} {\Omega_{\rm m}}\right ]^{1/3} - 1 \, .
\end{equation}
and one easily finds that, while the Hubble parameter remains finite,
\begin{equation}
\frac{H^2(z_s)}{H_0^2} = \Omega_\ell - \Omega_{\Lambda_{\rm b}} \, ,
\end{equation}
the deceleration parameter becomes singular as $z_s$ is approached. The difference
between the relatively mild {\em quiescent\/} singularities and the more spectacular {\em
Big Rip\/} singularities of phantom cosmology \cite{caldwell02} should be noted.
The latter are much more violent since the density, pressure and all derivatives of the
Hubble parameter diverge at the {\em Big Rip}; see also \cite{sing}.

\section{\bf Transient Acceleration on the Brane}

Setting $\Omega_\sigma = - 2 \sqrt{\Omega_\ell\, \Omega_{\Lambda_{\rm b}}}$ in
(\ref{hubble1}) leads to transient acceleration: the current acceleration of the universe
is a {\em transient phenomenon\/} sandwiched between two matter-dominated regimes
\cite{ss02}.

\section{\bf Cosmic mimicry} \label{mimicry}

The braneworld model in (\ref{hubble1}) has yet another remarkable property. For large
values of the brane tension $\Omega_\sigma$ and the (bulk) cosmological constant
$\Omega_{\Lambda_{\rm b}}$, and at redshifts lower than the {\em mimicry\/} redshift
\beq
(1+z_{\m})^3 = \frac{\Omega_\m \left(1 + \Omega_{\Lambda_{\rm b}} \right)}
{\left(\Omega_\m^{\rm LCDM} \right)^2} \, , \label{eq:new1}
\eeq
the expansion rate (\ref{hubble1}) on the brane reduces
to that in LCDM  \cite{mimicry}
\beq
\frac{H^2(z)}{H_0^2} \simeq \Omega^{\rm LCDM}_{\rm m} (1\!+\!z)^3 + 1 - \Omega^{\rm
LCDM}_{\rm m}\, , \label{eq:lcdm}
\eeq
where
\beq \label{lcdm}
\Omega^{\rm LCDM}_{\rm m} = \frac{\alpha}{\alpha \mp 1}\, \Omega_{\rm m}  \, , \qquad
\alpha = \frac{\sqrt{1 + \Omega_{\Lambda_\b}}}{\sqrt{\Omega_\ell}} \, .
\eeq
%\begin{equation} \label{alpha}
%\alpha = \frac{\sqrt{1 + \Omega_{\Lambda_\b}}}{\sqrt{\Omega_\ell}} \, .
%\end{equation}

\begin{figure*}[ht]
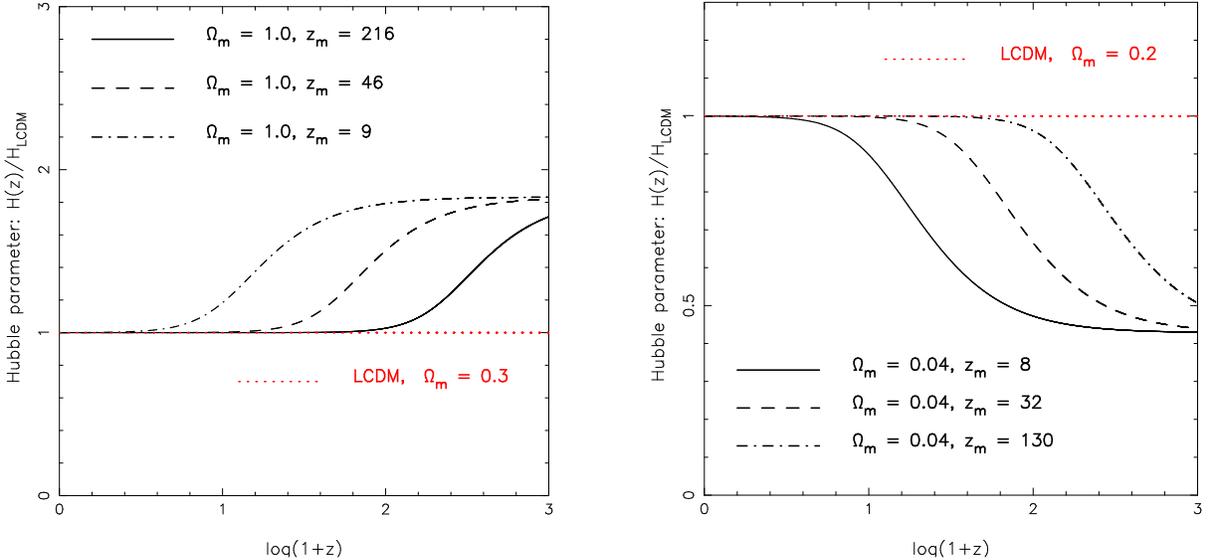

\centering
\begin{center}
$\begin{array}{@{\hspace{0.0in}}c@{\hspace{0.5in}}c}
\multicolumn{1}{l}{\mbox{}} &
\multicolumn{1}{l}{\mbox{}} \\ [-0.20in]
\epsfxsize=3.0in
\psfig{figure=b1.ps,width=0.45\textwidth,angle=-90} &
%\epsffile{b1.ps} &
\epsfxsize=3.0in
\psfig{figure=b2.ps,width=0.45\textwidth,angle=-90} \\
%\epsffile{b2.ps} \\
\end{array}$
\end{center}
\vspace{0.0cm} \caption{\small The {\em left panel\/} illustrates cosmic mimicry for the
Brane\,1 model. The Hubble parameter in three {\em high-density\/} Brane\,1 models with
$\Omega_\m = 1$ is shown. Also shown is the Hubble parameter in the LCDM model (red
dotted line) which closely mimics this braneworld but has a lower mass density
$\Omega_\m^{\rm LCDM} = 0.3$ ($\Omega_\l = 0.7$). The {\em right panel\/} illustrates
cosmic mimicry for the Brane\,2 model. Figure courtesy of  \cite{mimicry}. }
\label{fig:mimicry}
\end{figure*}

This property is dubbed {\em cosmic mimicry\/} for the following reasons:

\begin{itemize}

\item A Brane\,1 model, which at high redshifts expands with density parameter
$\Omega_\m$, at lower redshifts {\em masquerades as a LCDM universe\/} with a
{\em smaller value\/} of the density parameter. In other words, at low
redshifts, the Brane\,1 universe expands as the LCDM model (\ref{eq:lcdm}) with
$\Omega^{\rm LCDM}_{\rm m} < \Omega_{\rm m}$ [where $\Omega^{\rm LCDM}_{\rm m}$
is determined by (\ref{lcdm}) with the lower (``$+$'') sign].

\item A Brane\,2 model at low redshifts also masquerades as LCDM but with a
{\em larger value\/} of the density parameter. In this case, $\Omega^{\rm
LCDM}_{\rm m} > \Omega_{\rm m}$ with $\Omega^{\rm LCDM}_{\rm m}$ being
determined by (\ref{lcdm}) with the upper (``$-$'') sign.

\end{itemize}

Cosmic mimicry is illustrated in figure~\ref{fig:mimicry}.

\section{\bf Loitering Braneworld}

\begin{figure*}[ht]
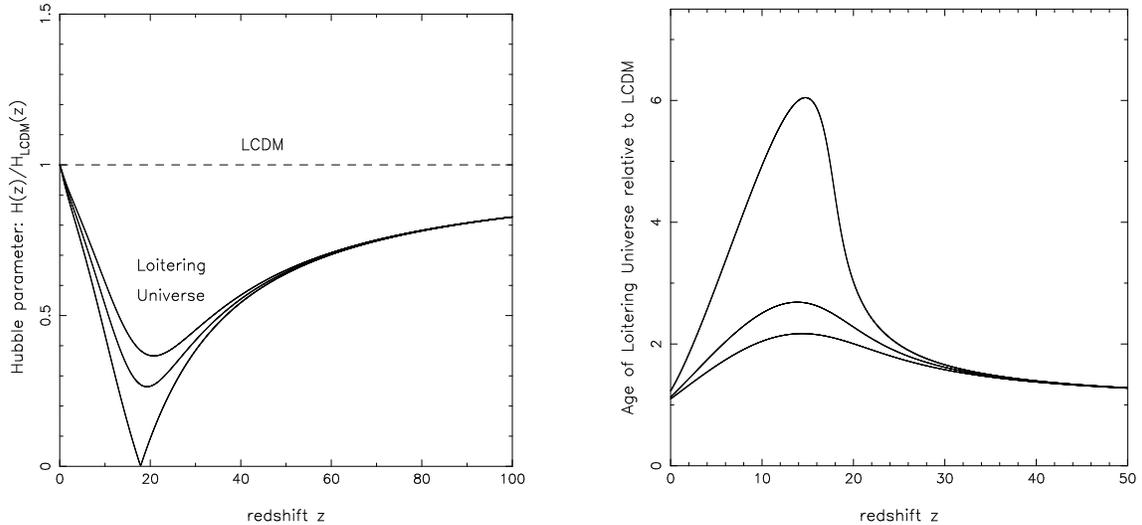

\centering
\begin{center}
$\begin{array}{@{\hspace{0.0in}}c@{\hspace{0.5in}}c} \multicolumn{1}{l}{\mbox{}} &
\multicolumn{1}{l}{\mbox{}} \\ [-0.20in]
\epsfxsize=2.7in \epsffile{loiter1.epsi} &
\epsfxsize=2.7in \epsffile{loiter2.epsi} \\
\end{array}$
\end{center}
\vspace{0.0cm}
\caption{\small
Left: Hubble parameter, with respect to LCDM,
 for three universes, all
loitering at $z_{\rm
loit} \simeq 18$.
Right: Ages of these loitering models relative to the age
of LCDM. The age increase arises because
$t(z) = \int_z^\infty \frac{dz'}{(1+z')H(z')}$;
a lower value of $H(z)$ clearly boosts the age of the
universe.
 Figure courtesy of  \cite{loiter}.
}
\label{fig:loiter}
\end{figure*}

An interesting aspect of the Braneworld models (\ref{hubble}), (\ref{eq:screening}) is
that they can {\em loiter\/} \cite{lemaitre,loiter}.  Loitering is characterized by the
fact that the Hubble parameter dips in value over a narrow redshift range referred to as
the `loitering epoch' \cite{sfs92}. In the model under consideration, it can occur even
in a spatially flat or open universe and is ensured by the presence of the dark-radiation
term $C/a^4$ in (\ref{eq:screening}).  During loitering, density perturbations are
expected to grow rapidly and, since the expansion of the universe slows down, its age
increases \cite{sfs92,loiter}. An epoch of loitering may, therefore, be expected to boost
the formation of high-redshift gravitationally bound systems including black holes and/or
Population III stars; see figure~\ref{fig:loiter}.

\end{document}